\documentclass[leqno, 10pt]{article}
\usepackage{times}
\usepackage{amsmath}
\usepackage{amssymb}
\usepackage{setspace}

\numberwithin{equation}{section}

\newtheorem{theorem}{Theorem}[section]
\newtheorem{definition}[theorem]{Definition}
\newtheorem{lemma}[theorem]{Lemma}
\newtheorem{corollary}[theorem]{Corollary}

\begin{document}

\date{}

\title{Combining intermediate propositional logics with classical logic}
\author{Steffen Lewitzka
\thanks{TR-PGCOMP-001/2015. Technical Report. Computer Science Graduate Program. Federal University of Bahia.
Departamento de Ci\^encia da Computa\c c\~ao,
UFBA,
40170-110 Salvador -- BA,
Brazil,
e-mail: steffen@dcc.ufba.br}}
\maketitle

\begin{abstract}
In \cite{lewjlc2}, we introduced a modal logic, called $L$, which combines intuitionistic propositional logic $IPC$ and classical propositional logic $CPC$ and is complete w.r.t. an algebraic semantics. However, $L$ seems to be too weak for Kripke-style semantics. In this paper, we add positive and negative introspection and show that the resulting logic $L5$ has a Kripke semantics. For intermediate logics $I$, we consider the parametrized versions $L5(I)$ of $L5$ where $IPC$ is replaced by $I$. $L5(I)$ can be seen as a classical modal logic for the reasoning about truth in $I$. From our results, we derive a simple method for determining algebraic and Kripke semantics for some specific intermediate logics. We discuss some examples which are of interest for Computer Science, namely the Logic of Here-and-There, G\"odel-Dummett Logic and Jankov Logic. Our method provides new proofs of completeness theorems due to Hosoi, Dummett/Horn and Jankov, respectively.  
\end{abstract}

\textbf{Keywords}: intuitionistic logic, intermediate logic, non-Fregean logic, Heyting algebra, Logic of Here-and-There, G\"odel-Dummett Logic, Jankov Logic

\section{Introduction}

The study of certain modal systems from the perspective of non-Fregean logic seems to be a promising approach (see e.g. \cite{lewjlc1, lewjlc2, lewsl, sus, blosus}). The basic classical non-Fregean logic is Suszko's Sentential Calculus with Identity $SCI$ \cite{blosus}. $SCI$ contains an identity connective $\equiv$ and extends classical propositional logic $CPC$ by the following identity axioms:\footnote{Instead of scheme (Id3), Suszko considers a collection of other axioms. However, it can be shown that that collection of axioms is equivalent with (Id3) modulo the rest (see \cite{lewjlc1}).}\\

\noindent (Id1) $\varphi\equiv\varphi$\\
(Id2) $(\varphi\equiv\psi)\rightarrow (\varphi\leftrightarrow\psi)$\\
(Id3) $(\varphi\equiv\psi)\rightarrow (\chi[x:=\varphi]\equiv\chi[x:=\psi])$\footnote{Formula $\chi[x:=\varphi]$ is the result of replacing every occurrence of variable $x$ in $\chi$ by $\varphi$.}\\

$\varphi\equiv\psi$ reads ``$\varphi$ and $\psi$ have the same meaning (denotation, \textit{Bedeutung})". While $(\varphi\equiv\psi)\rightarrow (\varphi\leftrightarrow\psi)$ is a theorem, its converse $(\varphi\leftrightarrow\psi)\rightarrow (\varphi\equiv\psi)$ is not. The latter says that $\varphi$ and $\psi$ have the same meaning whenever they have the same truth value. This is essentially what Suszko called the \textit{Fregean Axiom}. Logics without Fregean Axiom are called non-Fregean logics. We regard the denotation of a formula as a \textit{proposition} and refer to the axioms (Id1)--(Id3) above as the \textit{axioms of propositional identity}. In particular, we refer to (Id3) as the Substitution Principle SP. It corresponds to a general ontological principle as part of Leibniz' law and is sometimes called in the literature the \textit{Indiscernibility of Identicals}: identical entities can be substituted by each other in all contexts. Note that the modal systems S1--S5, introduced by C. Lewis as logics of \textit{strict implication}, satisfy (Id1)--(Id2) if we define propositional identity $\varphi\equiv\psi$ as \textit{strict equivalence} $\square(\varphi\rightarrow\psi)\wedge\square(\psi\rightarrow\varphi)$. In \cite{lewjlc1, lewsl} we saw that under this assumption, S3 is the weakest Lewis modal logic which also satisfies SP, i.e. identity axiom (Id3). There is no known intuitive semantics for Lewis system S1. Nevertheless, in \cite{lewjlc1} we were able to present an algebraic, non-Fregean-style semantics for the slightly stronger system S1+SP which results from S1 by adding all formulas of the form SP as theorems. Thus, the ``Lewis-style" modal logics S1+SP, S3, S4, S5 can be viewed as specific $SCI$-theories where propositional identity $\varphi\equiv\psi$ is given as strict equivalence $\square(\varphi\leftrightarrow\psi)$. These observations led to the development of logic $L$ \cite{lewjlc2} which extends the intuitionistic version of S1+SP by an axiom for a \textit{disjunction property} and the theorem \textit{tertium non datur}. It turns out that $L$ is a conservative extension of $CPC$ and contains a copy of intuitionistic propositional logic $IPC$ by means of the embedding $\varphi\mapsto\square\varphi$ from $IPC$ to $L$. That is, $L$ is a modal logic that combines $IPC$ and $CPC$. $L$ has a non-Fregean-style semantics given by a class of specific Heyting algebras with a modal operator and a designated ultrafilter. However, we are not able to provide a Kripke-style semantics for $L$. 

This paper is organized as follows. First, we present axiomatization and algebraic semantics of $L5$, the logic which results from $L$ by adding the axioms of positive and negative introspection. Many facts concerning logic $L$ can be adopted. In the following sections, we introduce Kripke-style semantics of $L5$ and prove equivalence to algebraic semantics by showing that each $L5$-model corresponds to a Kripke frame which satisfies exactly the same formulas, and vice-versa. Next, we generalize the approach and study the parametrized versions of $L5$: for each intermediate logic $I$, we consider modal logic $L5(I)$ which extends $L5$ in the sense that $IPC$ is replaced by $I$. The main result from \cite{lewjlc2} then can be re-formulated in the following way: $\varphi$ derives from $\Phi$ in logic $I$ iff $\square\varphi$ derives from $\square\Phi$ in $L5(I)$, for propositional $\Phi\cup\{\varphi\}$. We conclude that modal logic $L5(I)$ is a conservative extension of $CPC$ and contains a copy of $I$ by means of the embedding $\varphi\rightarrow\square\varphi$ from $I$ to $L5(I)$. That is, $L5(I)$ combines $I$ and $CPC$ in a similar way as $L$ combines $IPC$ and $CPC$. Our results give rise to a simple method of deriving the algebraic and Kripke-style semantics of some specific intermediate logics. Discussing the particular cases of the Logic of Here-and-There, G\"odel-Dummett Logic and Jankov Logic, we are able to establish new proofs, with some simplifications, of corresponding completeness results found in the literature.

\section{Modal logic $L5$}

The language of modal propositional logic is inductively defined in the usual way over a set of variables $x_0, x_1, ... $, logical connectives $\wedge$, $\vee$, $\rightarrow$, $\bot$ and the modal operator $\square$. $Fm$ denotes the set of formulas, and $Fm_0\subseteq Fm$ denotes the set of \textit{propositional formulas}, i.e. formulas without modal operator $\square$. We use the following abbreviations:\\

\noindent $\neg\varphi :=\varphi\rightarrow\bot$\\
$\top :=\neg\bot$\\
$\varphi\leftrightarrow\psi :=(\varphi\rightarrow\psi)\wedge (\psi\rightarrow\varphi)$\\
$\varphi\equiv\psi := \square(\varphi\rightarrow\psi)\wedge\square(\psi\rightarrow\varphi)$ (``propositional identity = strict implication") \\
$\square\Phi :=\{\square\psi\mid\psi\in\Phi\}$\\

We consider the following axiom  schemes:\\

\noindent (i) theorems of $IPC$\footnote{We mean all formulas which have the form of some IPC-theorem. For instance, $\square x\rightarrow \square x$ has the form $\varphi\rightarrow\varphi$ and is therefore an axiom.}\\
(ii) $\square\varphi\rightarrow\varphi$\\
(iii) $\square(\varphi\rightarrow\psi)\rightarrow(\square(\psi\rightarrow\chi)\rightarrow\square(\varphi\rightarrow\chi))$\\
(iv) $\square(\varphi\vee\psi)\rightarrow(\square\varphi\vee\square\psi)$ \\
(v) $\square\varphi\rightarrow\square\square\varphi$ \\
(vi) $\neg\square\varphi\rightarrow\square\neg\square\varphi$  \\

We call scheme (iv) the disjunction property. Schemes (v) and (vi) are the axioms of positive and negative introspection, respectively.
The inference rules are Modus Ponens (MP) and Axiom Necessitation (AN) ``If $\varphi$ is an axiom, then infer $\square\varphi$." Furthermore, we add formulas of the form SP, i.e. (Id3) above, and \textit{tertium non datur} $\varphi\vee\neg\varphi$ as theorems. Note that rule AN only applies to axioms, i.e. formulas of the form (i)--(vi). We call the resulting deductive system $L5$ and write $\Phi\vdash_{L5}\varphi$ if formula $\varphi$ is derivable from $\Phi$ in $L5$. Recall that $L$ \cite{lewjlc2} is $L5$ minus (v) and (vi). Also observe that we obtain Lewis modal logic S1 if we drop (iv)--(vi), replace $IPC$ by $CPC$ in (i), and replace scheme SP by the weaker rule of Substitutions of Proved Strict Equivalents (SPSE) ``If $\varphi\equiv\psi$ is a theorem, then $\chi[x:=\varphi]\equiv\chi[x:=\psi]$ is a theorem" (see e.g. \cite{hugcres} for a discussion about Lewis modal systems). In logic $L5$ as well as in $L$, the modal operator, if restricted to propositional formulas, can be seen as a predicate for provability (= intuitionistic truth). The axioms (ii)--(vi) then express principles of constructive logic. For instance, scheme (iv) says that the existence of a proof of $\varphi\vee\psi$ implies the existence of a proof of $\varphi$ or a proof of $\psi$.\footnote{The converse $(\square\varphi\vee\square\psi)\rightarrow\square(\varphi\vee\psi)$ is derivable.} This constructive principle cannot be expressed in $IPC$ itself.\\

As in logic $L$, the Deduction Theorem holds. Interestingly, the modal laws
\begin{itemize}
\item $\square\varphi\leftrightarrow (\varphi\equiv \top)$ (``There is exactly one necessary proposition.")
\item $\square(\varphi\rightarrow\psi)\rightarrow(\square\varphi\rightarrow\square\psi)$
\item $\square(\varphi\wedge\psi)\leftrightarrow(\square\varphi\wedge\square\psi)$
\item $(\square(\varphi\rightarrow\psi)\wedge\square(\psi\rightarrow\varphi))\leftrightarrow\square(\varphi\leftrightarrow\psi)$
\end{itemize}
are theorems. Derivations of the first two theorems can be found in \cite{lewjlc1}. The third theorem derives similarly as in normal modal logics using modal law K (i.e. the second theorem). Finally, the last theorem is a consequence of the third one. In particular, propositional identity $\varphi\equiv\psi$ is given by $\square(\varphi\leftrightarrow\psi)$.

\section{Denotational semantics} 

In \cite{lewjlc2}, we presented an algebraic, non-Fregean-style, semantics for logic $L$. We also use the term \textit{denotational semantics} because there is an explicitly given function that maps formulas to their denotations/meanings as elements of a model-theoretic universe. A model for $L$, which we call here a $L$-model, is a Heyting algebra 
\begin{equation*}
\mathcal{M}=(M,\mathit{TRUE},f_\bot, f_\top, f_\rightarrow,f_\vee,f_\wedge,f_\square)
\end{equation*}
with a designated ultrafilter $\mathit{TRUE}\subseteq M$ on universe $M$ and an operation $f_\square$ such that for all $m,m',m''\in M$ the following truth conditions are fulfilled ($\le$ is the lattice ordering):
\begin{enumerate}
\item $f_\square(m)\le m$
\item $f_\square(f_\rightarrow(m,m'))\le f_\rightarrow(f_\square(f_\rightarrow(m',m'')),f_\square(f_\rightarrow(m,m'')))$
\item $f_\square(f_\vee(m,m'))\le f_\vee(f_\square(m),f_\square(m'))$
\item $f_\square(m)\in\mathit{TRUE}\Leftrightarrow m=f_\top$
\end{enumerate}
We regard $M$ as a \textit{propositional universe} being $\mathit{TRUE}\subseteq M$ the set of (classically) \textit{true propositions}. $f_\top$, $f_\bot$ are the top and the bottom element of the Heyting algebra and stand for intuitionistic truth and falsity, respectively.\footnote{Note that we do not regard the elements of the underlying Heyting algebra as ``generalized truth values" as it is sometimes the case in the literature when algebraic semantics of $IPC$ is discussed.}\\
 
An important feature of a $L$-model is the \textit{disjunction property} DP: for all $m,m'\in M$, $f_\vee(m,m')=f_\top$ iff $m=f_\top$ or $m'=f_\top$. Note that DP follows from truth conditions (iii) and (iv) of a $L$-model and is not a general property of Heyting algebras. That is, DP defines a specific subclass of Heyting algebras.\\ 

A $L5$-model is a $L$-model satisfying the following additional truth condition:

(v) For all $m\in M$, 
\begin{equation*}
\begin{split}
f_\square(m)=
\begin{cases}
&f_\top,\text{ if }m=f_\top\\
&f_\bot,\text{ else.}
\end{cases}
\end{split}
\end{equation*}

Note that truth condition (v) ensures soundness of the axioms of positive and negative introspection if we consider the definition of satisfaction below. Also observe that in a $ L5$-model, the truth conditions (i), (ii)  and (iv) follow already from truth condition (v).\\

Given a model $\mathcal{M}$, an assignment in $\mathcal{M}$ is a function $\gamma\colon V\rightarrow M$ that extends in the canonical way to a function on $Fm$, i.e. $\gamma(\bot)=f_\bot$, $\gamma(\top)=f_\top$, $\gamma(\square\varphi)=f_\square(\gamma(\varphi))$, $\gamma(\varphi*\psi)=f_*(\gamma(\varphi),\gamma(\psi))$, for $*\in\{\vee,\wedge,\rightarrow\}$. A $L5$-interpretation is a tuple $(\mathcal{M},\gamma)$ consisting of a $L5$-model and a corresponding assignment. The relation of satisfaction is defined by $(\mathcal{M},\gamma)\vDash\varphi :\Leftrightarrow \gamma(\varphi)\in\mathit{TRUE}$ and extends in the usual way to sets of formulas. Finally, the relation of logical consequence in logic $L5$ is defined by $\Phi\Vdash_{L5}\varphi :\Leftrightarrow$ $(\mathcal{M},\gamma)\vDash\Phi$ implies $(\mathcal{M},\gamma)\vDash\varphi$, for every $L5$-interpretation $(\mathcal{M},\gamma)$.\\

The following is not hard to prove (see e.g. \cite{lewjlc1}): 
\begin{equation}\label{500}
(\mathcal{M},\gamma)\vDash\varphi\equiv\psi\Leftrightarrow\gamma(\varphi)=\gamma(\psi).
\end{equation}
That is, $\varphi\equiv\psi$ is true iff $\varphi$ and $\psi$ denote the same proposition. This is precisely the intended meaning of an identity connective in a denotational semantics, and that's why we refer to it as \textit{propositional identity}.\footnote{When we say that a formula $\varphi$ \textit{denotes} a proposition $m\in M$ of a given model $\mathcal{M}$, then we are assuming a given assignment $\gamma$ with $\gamma(\varphi)=m$.}\\

If $(\mathcal{M},\gamma)$ is any interpretation and $\varphi\leftrightarrow\psi$ is any theorem of $CPC$, such as $\neg\neg\chi\leftrightarrow\chi$, then $(\mathcal{M},\gamma)\vDash\varphi\leftrightarrow\psi$ but not necessarily $(\mathcal{M},\gamma)\vDash\varphi\equiv\psi$. That is, Fregean Axiom $(\varphi\leftrightarrow\psi)\rightarrow (\varphi\equiv\psi)$ does not hold. This is in the very spirit of non-Fregean logic: two formulas with the same truth value may have distinct denotations (meanings). If we consider the preorder defined by $m\preceq m' :\Leftrightarrow f_\rightarrow(m,m')\in \mathit{TRUE}$, then the underlying Heyting algebra is a Boolean prealgebra with preorder $\preceq$, according to Definition 3.1 in \cite{lewsl}.\footnote{Roughly speaking, a Boolean prealgebra is a structure that generalizes a Boolean algebra in the sense that the underlying lattice ordering is no longer a partial ordering but a preorder, i.e., the axiom of antisymmetry is not necessarily satisfied.} In fact, the quotient algebra of the underlying Heyting algebra modulo ultrafilter $\mathit{TRUE}$ is the two-element Boolean algebra with $\mathit{TRUE}$ as top element. We proved in \cite{lewsl} that Boolean prealgebras and models of basic non-Fregean logic $SCI$ are essentially the same mathematical objects.\\ 

We call an interpretation $(\mathcal{M},\gamma)$ surjective if $\gamma\colon Fm\rightarrow M$ is surjective, i.e. if for each $m\in M$ there is a $\varphi\in Fm$ such that $\gamma(\varphi)=m$. Note that for any interpretation $(\mathcal{M},\gamma)$, the set $\gamma(Fm)\subseteq M$ is the universe $M'$ of a submodel $\mathcal{M'}$ of $\mathcal{M}$ in the sense that the operations of $\mathcal{M}$ restricted to $M'=\gamma(Fm)$ form a Heyting algebra that satisfies the truth conditions of a model $\mathcal{M'}$. In fact, if $m,m'\in M'$ then $f_\wedge(m,m')=f_\wedge(\gamma(\varphi),\gamma(\psi))=\gamma(\varphi\wedge\psi)\in M'$, for some $\varphi,\psi\in Fm$, and similarly for the remaining operations. Then it is clear that $M'$ forms a Heyting algebra. It is also clear that the truth conditions of a model hold for all subsets of the universe, particularly for $M'$. Furthermore, one easily recognizes that $\mathit{TRUE}\cap\gamma(Fm)$ is an ultrafilter on $M'$. Thus, $(\mathcal{M},\gamma)\vDash\varphi\Leftrightarrow\gamma(\varphi)\in \mathit{TRUE}\Leftrightarrow\gamma(\varphi)\in\mathit{TRUE}\cap\gamma(Fm)\Leftrightarrow (\mathcal{M'},\gamma)\vDash\varphi$. That is, the interpretations $(\mathcal{M},\gamma)$ and $(\mathcal{M'},\gamma)$ satisfy exactly the same formulas. Therefore, we may assume in the following that all interpretations are surjective. \\

The completeness proof for $L$ \cite{lewjlc2} extends straightforwardly to the case of logic $L5$ and the corresponding class of $L5$-models:

\begin{theorem}\label{170}
Logic L5 is sound and complete w.r.t. the class of all L5-models. That is, for any set of formulas $\Phi\cup\{\varphi\}$, $\Phi\vdash_{L5}\varphi\Leftrightarrow\Phi\Vdash_{L5}\varphi$.
\end{theorem}

\section{Kripke-style semantics for Logic $L5$}

We were unable to find a Kripke semantics for logic $L$. The addition of axiom schemes for positive and negative introspection (schemes (v) and (vi)) to $L$ enables us to establish a natural Kripke-style semantics for the resulting logic $L5$. A $L5$-frame $(W,R)$ is given by a non-empty set $W$ of worlds and a partial ordering $R\subseteq W\times W$, called accessibility relation, with the property that there is a $R$-smallest element, which we usually denote by $w_B$ (the bottom of the frame), and every $R$-chain has an upper bound in $W$. Note that Zorn's Lemma implies that each $w\in W$ accesses a $R$-maximal element. An assignment in a given $L5$-frame $(W,R)$ is a function $g\colon V\rightarrow Pow(W)$ satisfying the following \textit{monotonicity condition}: For all $w,w'\in W$ and $x\in V$, if $wRw'$ and $w\in g(x)$, then $w'\in g(x)$. The satisfaction relation is defined as follows. Suppose $(W,R)$ is a $L5$-frame, $g$ is an assignment in $(W,R)$, and $w\in W$. Then\\

\noindent $(w,g)\nvDash \bot$\\
$(w,g)\vDash x :\Leftrightarrow w\in g(x)$\\
$(w,g)\vDash \varphi\vee\psi :\Leftrightarrow (w,g)\vDash\varphi$ or $(w,g)\vDash\psi$\\
$(w,g)\vDash \varphi\wedge\psi :\Leftrightarrow (w,g)\vDash\varphi$ and $(w,g)\vDash\psi$\\
$(w,g)\vDash \varphi\rightarrow\psi :\Leftrightarrow$ for all $w'\in W$ with $wRw'$, $(w',g)\vDash\varphi$ implies $(w',g)\vDash\psi$\\
$(w,g)\vDash \square\varphi :\Leftrightarrow (w_B,g)\vDash\varphi$\\

Note that semantics of logical connectives is defined as in usual intuitionistic Kripke models. The next monotonicity result, which also holds in $IPC$, can be shown by induction on formulas.

\begin{lemma}\label{180}
If $(W,R)$ is a $L5$-frame and $g\in Pow(W)^V$ is an assignment, then for all $w,w'\in W$ and all formulas $\varphi$: if $(w,g)\vDash\varphi$ and $wRw'$, then $(w',g)\vDash\varphi$.
\end{lemma}

\begin{lemma}\label{190}
Let $(W,R)$ be a $L5$-frame, $w\in W$ and $g$ an assignment. Then for any formula $\varphi$,
\begin{itemize}
\item $(w,g)\vDash\square\varphi\rightarrow\square\square\varphi$
\item $(w,g)\vDash\neg\square\varphi\rightarrow\square\neg\square\varphi$.
\end{itemize}
\end{lemma}

\paragraph*{Proof.}
We leave the first claim as an exercise and outline the proof of the second statement. $(w,g)\vDash\neg\square\varphi$ means that $(w',g)\nvDash\square\varphi$, for all $w'\in W$ with $wRw'$. This implies $(w,g)\vDash\neg\square\varphi$ implies $(w_B,g)\nvDash\varphi$ implies $(w',g)\nvDash\square\varphi$, for all $w'\in W$, implies $(w_B,g)\vDash\neg\square\varphi$ implies $(w,g)\vDash\square\neg\square\varphi$. Now, the claim follows. Q.E.D.

\section{Translation results}

Of course, we expect that our algebraic and Kripke-style semantics for logic $L5$ are equivalent in the sense that both lead to consequence relations which model precisely the relation $\vdash_{L5}$ of derivability. Instead of proving completeness of $L5$ w.r.t. Kripke semantics directly, we show in this section in which way algebraic and Kripke semantics translate into each other. The following basic facts about filters in Heyting algebras, possibly known to the reader, will be useful.

\begin{lemma}\label{200}
Let $\mathcal{H}$ be a Heyting algebra. Then:\\
(a) Every filter is the intersection of a set of prime filters.\\
(b) Let $m_1,m_2\in H$ and $P$ be a prime filter. If for all prime filters $P'\supseteq P$, $m_1\in P'$ implies $m_2\in P'$, then $f_\rightarrow(m_1,m_2)\in P$. \\
(c) If $U$ is an ultrafilter, then for all $m,m'\in H$:
\begin{itemize}
\item $m\in U$ or $f_\neg(m):=f_\rightarrow(m,f_\bot)\in U$
\item $f_\rightarrow(m,m')\in U$ iff [$m\notin U$ or $m'\in U$] iff $f_\vee(f_\neg(m),m')\in U$
\item $U$ is a prime filter.
\end{itemize}
\end{lemma}

\paragraph*{Proof.}
(a): Let $F$ be a filter, and let $X$ be the set of prime filters containing $F$. Since every filter is contained in an ultrafilter which, by the last statement of the Lemma, is a prime filter, $X$ is non-empty. Obviously, $F\subseteq \bigcap X$. Suppose there is $m\in \bigcap X\smallsetminus F$. By a standard application of Zorn's Lemma, we derive the existence of an ultrafilter $U$ that contains $F$ but not $m$. Then $U\in X$. This contradicts the hypothesis $m\in\bigcap X$. Thus, $\bigcap X=F$.\\
(b): Let $m_1,m_2\in H$ and $P$ be a prime filter. We consider the quotient Heyting algebra $\mathcal{H}'$ of $\mathcal{H}$ modulo $P$. That is, the elements of $\mathcal{H'}$ are the equivalence classes $\overline{m}$ of $m\in M$ modulo the equivalence relation $\sim$ defined by $m\sim m'\Leftrightarrow$ [$f_\rightarrow(m,m')\in P$ and $f_\rightarrow(m',m)\in P$]. Then one easily checks that $P$ is the equivalence class of $f_\top$ modulo $\sim$, and it is the top element $f'_\top$ of $\mathcal{H'}$.\\
\textbf{Claim1}: Let $m,m'\in H$. If $\overline{m}\in F'$ implies $\overline{m'}\in F'$, for all filters $F'$ of $\mathcal{H}'$, then $\overline{m}\le' \overline{m'}$, where $\le'$ is the lattice ordering of $\mathcal{H}'$.\\
\textit{Proof of Claim1}. Suppose $\overline{m}\nleqslant' \overline{m'}$. Consider the filter $G=\{\overline{m''}\mid \overline{m}\le' \overline{m''}\}$. Then $\overline{m}\in G$ and $\overline{m'}\notin G$. We have proved the Claim.\\
\textbf{Claim2}: Let $m,m'\in H$. If $\overline{m}\in F'$ implies $\overline{m'}\in F'$, for all prime filters $F'$ of $\mathcal{H}'$, then $\overline{m}\le' \overline{m'}$, where $\le'$ is the lattice ordering of $\mathcal{H}'$.\\
\textit{Proof of Claim2}. Claim2 follows from Claim1 together with (a).\\
\textbf{Claim3}: If $F'$ is a (prime) filter of $\mathcal{H}'$, then $F=\{m\mid\overline{m}\in F'\}$ is a (prime) filter of $\mathcal{H}$ extending $P$.\\
\textit{Proof of Claim3}. Suppose $m\in F$ and $m\le m'$. Then $f_\rightarrow(m,m')=f_\top$. Thus, $\overline{f_\rightarrow(m,m')}=P=f'_\top$. That is, $f'_\rightarrow(\overline{m},\overline{m'})=f'_\top$ and therefore $\overline{m}\le'\overline{m'}$. It follows that $\overline{m'}\in F'$ and $m'\in F$. The remaining filter properties follow straightforwardly. $m\in P$ implies $\overline{m}=P=f'_\top\in F'$ implies $m\in F$. Thus, $P\subseteq F$ and Claim3 holds true.\\
Now suppose the premises of (b) are true. Let $F'$ be any prime filter of $\mathcal{H}'$ and $\overline{m_1}\in F'$. Then, by Claim3, $m_1\in F=\{m\mid\overline{m}\in F'\}$ and $F$ is a prime filter of $\mathcal{H}$ with $P\subseteq F$. By hypothesis of (b), $m_2\in F$. Thus, $\overline{m_2}\in F'$. By Claim2, $\overline{m_1}\le' \overline{m_2}$. Then $\overline{f_\rightarrow(m_1,m_2)}=f'_\top=P$. That is, $f_\rightarrow(m_1,m_2)\in P$.\\
(c) It is not hard to check that the quotient algebra of $\mathcal{H}$ modulo ultrafilter $U$ is the two-element Boolean algebra with top element $\overline{f_\top}=U$. Alternatively, one can show that the map $h\colon H\rightarrow\{\overline{f_\bot},\overline{f_\top}\}$, defined by $h(m)= \overline{f_\top} :\Leftrightarrow m\in U$, is an homomorphism of Heyting algebras. The assertions of the Lemma then follow by switching between the elements of $\mathcal{H}$ and their corresponding congruence classes $\overline{f_\top}$ and $\overline{f_\bot}$, i.e. the two elements of the quotient algebra. Q.E.D.\\

There is a close connection between Heyting algebras and intuitionistic Kripke frames which can be studied under different aspects (see e.g. \cite{fit, bezjon}). The next two Theorems give an approach from the perspective of our semantical investigations. The construction developed in the proof of Theorem \ref{270} (cf. [Theorem 6.1 \cite{lewsl}]) will be particularly useful for the method of determining Kripke semantics of some intermediate logics, as discussed in the last section.

\begin{theorem}\label{250}
Suppose $\mathcal{M}$ is a $L5$-model and $\gamma\in M^V$ is an assignment. Then there are a $L5$-frame $(W,R)$, a maximal world $w_T\in W$ and an assignment $g\in Pow(W)^V$ such that for all formulas $\varphi$:
\begin{equation*}
(\mathcal{M},\gamma)\vDash\varphi\Leftrightarrow (w_T,g)\vDash\varphi.
\end{equation*}
\end{theorem}

\paragraph*{Proof.}
Let $W$ be the set of all prime filters of the underlying Heyting algebra on $\mathcal{M}$. Then $\mathit{TRUE}$ is a maximal element of $W$ and $w_B:=\{f_\top\}$ is the bottom world w.r.t. the accessibility relation $R$ which is given by set inclusion: $wRw' :\Leftrightarrow w\subseteq w'$. The union of a chain of prime filters is again a prime filter. Thus, every chain in $W$ has an upper bound in $W$ and $(W,R)$ fulfills the requirements of a $L5$-frame. For a given assignment $\beta\in M^V$, define the function $g_\beta\colon V\rightarrow Pow(W)$ by $x\mapsto \{w\in W\mid \beta(x)\in w\}$. Then $wRw'$ together with $w\in g_\beta(x)$ implies $w'\in g_\beta(x)$. That is, function $g_\beta$ fulfills the monotonicity condition and is in fact an assignment in $(W,R)$.\\
\textbf{Claim}: Let $\beta\in M^V$ be any assignment in model $\mathcal{M}$. Then for all $w\in W$: 
\begin{equation*}
(w,g_\beta)\vDash\varphi\Leftrightarrow\beta(\varphi)\in w.
\end{equation*}
We prove the Claim by induction on $\varphi$, simultaneously for all $w\in W$. In the basis case $\varphi=x\in V$, the Claim follows from the definition of assignment $g_\beta$. Let $\varphi=\psi\vee\chi$. Then 
\begin{equation*}
\begin{split}
(w,g_\beta)\vDash\psi\vee\chi &\Leftrightarrow (w,g_\beta)\vDash\psi\text{ or }(w,g_\beta)\vDash\chi\\
&\Leftrightarrow\beta(\psi)\in w\text{ or }\beta(\chi)\in w,\text{ by induction hypothesis}\\
&\Leftrightarrow f_\vee(\beta(\psi),\beta(\psi))\in w,\text{ since }w\text{ is a prime filter} \\
&\Leftrightarrow \beta(\varphi\vee\psi)\in w,\text{ by definition of an assignment}
\end{split}
\end{equation*}
The case $\varphi=\psi\wedge\chi$ follows similarly. Suppose $\varphi=\psi\rightarrow\chi$. Then, again by induction hypothesis, we get
\begin{equation*}
\begin{split}
(w,g_\beta)\vDash\psi\rightarrow\chi &\Leftrightarrow (w',g_\beta)\vDash\psi\text{ implies }(w',g_\beta)\vDash\chi,\text{ for each }w'\text{ with }wRw'\\
&\Leftrightarrow\beta(\psi)\in w'\text{ implies }\beta(\chi)\in w',\text{ for each }w'\text{ with }wRw'\\
&\overset{*}{\Leftrightarrow}f_\rightarrow(\beta(\psi),\beta(\chi))\in w\\
&\Leftrightarrow\beta(\psi\rightarrow\chi)\in w
\end{split}
\end{equation*}
The left-to-right direction of (*) follows from Lemma \ref{200} (b). The right-to-left direction of (*) follows from the fact that $f_\rightarrow(m,m')$ is the relative pseudo-complement of $m$ w.r.t. $m'$ in the underlying lattice.\\
Finally, let $\varphi=\square\psi$. Then
\begin{equation*}
\begin{split}
(w,g_\beta)\vDash\square\psi &\Leftrightarrow (w_B,g_\beta)\vDash\psi\\
&\Leftrightarrow \beta(\psi)=f_\top,\text{ by induction hypothesis and the definition of }w_B\\
&\Leftrightarrow \beta(\square\psi)=f_\square(\beta(\psi))=f_\top,\text{ by truth conditions of a $L5$-model}
\end{split}
\end{equation*}
In particular, $(w,g_\beta)\vDash\square\psi\Rightarrow \beta(\square\psi)\in w$. On the other hand, $\beta(\square\psi)\in w$ implies $\beta(\square\psi)=f_\square(\beta(\psi))\neq f_\bot$ because $w$ is a filter and does not contain $f_\bot$. By truth condition (v),  $f_\square(\beta(\psi))=f_\top$ and $\beta(\psi)=f_\top$. By the equivalences above, this implies $(w,g_\beta)\vDash\square\psi$. Hence, the Claim holds true. Then for the world $w_T=\mathit{TRUE}\in W$ we have: 
\begin{equation*}
(w_T,g_\gamma)\vDash\varphi\overset{Claim}{\Longleftrightarrow}\gamma(\varphi)\in\mathit{TRUE}\Longleftrightarrow(\mathcal{M},\gamma)\vDash\varphi,
\end{equation*}
for any formula $\varphi$. Q.E.D.\\

Observe that truth condition (v) of a $L5$-model is crucial for the last case of the induction step in the above proof. The proof does not work with any $L$-model not satisfying truth condition (v).

\begin{theorem}\label{270}
Let $(W,R)$ be a $L5$-frame, $g\colon V\rightarrow Pow(W)$ an assignment and $w_T\in W$ a maximal element of $W$. Then there are a $L5$-model $\mathcal{M}$ and an assignment $\gamma\colon V\rightarrow M$ such that for all formulas $\varphi$:
\begin{equation*}
(\mathcal{M},\gamma)\vDash\varphi\Leftrightarrow (w_T,g)\vDash\varphi.
\end{equation*}
\end{theorem}

\paragraph*{Proof.}
Suppose we are given a $L5$-frame $(W,R)$ with a maximal world $w_T\in W$ and an assignment $g\in Pow(W)^V$. We define an equivalence relation $\approx$ on the set of formulas by
\begin{equation*}
\varphi\approx\psi :\Leftrightarrow (w_B,g)\vDash\varphi\leftrightarrow\psi,
\end{equation*}
where $w_B$ is the bottom world. Thus, $\varphi\approx\psi$ iff $(w,g)\vDash\varphi\equiv\psi$, for any world $w\in W$. One easily checks that $\approx$ respects the logical connectives as well as the modal operator. Thus, $\approx$ is a congruence relation on the set of formulas. By $\overline{\varphi}$ we denote the congruence class of a formula $\varphi$ modulo $\approx$. Then we define the ingredients of our $L5$-model by $M:=\{\overline{\varphi}\mid\varphi\in Fm\}$, $\mathit{TRUE}:=\{\overline{\varphi}\mid(w_T,g)\vDash\varphi\}$, $f_\top:=\overline{\top}$, $f_\bot:=\overline{\bot}$, $f_\square(\overline{\varphi}):=\overline{\square\varphi}$ and $f_{*}(\overline{\varphi},\overline{\psi}):=\overline{\varphi * \psi}$, for $*\in\{\vee,\wedge,\rightarrow\}$. Since $\approx$ is a congruence relation, all these ingredients are well-defined. We must show that $\mathcal{M}=(M,\mathit{TRUE},f_\bot, f_\top, f_\rightarrow,f_\vee,f_\wedge,f_\square)$ fulfills the conditions of a $L5$-model.\\
\textbf{Claim}: $(M,f_\bot, f_\top, f_\rightarrow,f_\vee,f_\wedge)$ is a Heyting algebra.\\
\textit{Proof of the Claim}. The class of Heyting algebras can be axiomatized by a set of equations which correspond to theorems of $IPC$ of the form $\varphi\leftrightarrow\psi$. Then at every world in every Kripke model of intuitionistic logic, $\varphi$ is true iff $\psi$ is true. Since our frames are in particular Kripke models of $IPC$, we get $\varphi\approx\psi$ for every theorem $\varphi\leftrightarrow\psi$ of $IPC$. Thus, $\overline{\varphi}=\overline{\psi}$ and we have a Heyting algebra. It is clear by the definitions that $\mathit{TRUE}$ is an ultrafilter on $M$.\\
It remains to show that $\mathcal{M}$ satisfies the truth conditions (i)--(v) of a $L5$-model. Recall that in any Heyting algebra: $f_\rightarrow(m,m')=f_\top$ iff $m\le m'$. Also, we observe that $\overline{\varphi}=\overline{\top}$ iff $\varphi\approx \top$ iff $(w_B,g)\vDash\varphi$, where $w_B$ is the bottom world. Then, in order to verify truth condition (i), it suffices to show that $(w_B,g)\vDash \square\varphi\rightarrow\varphi$. This obviously holds true. Similarly, one checks truth conditions (ii) and (iii). Finally, we check truth conditions (iv) and (v). 
On the one hand, $\overline{\varphi}=f_\top$ implies $(w_B,g)\vDash\varphi$ implies $(w_B,g)\vDash\square\varphi$ implies $f_\square(\overline{\varphi})=f_\top$. On the other hand, $\overline{\varphi}\neq f_\top$ implies $(w',g)\nvDash\varphi$, for some $w'\in W$, implies $(w_B,g)\nvDash\varphi$ implies $(w,g)\nvDash\square\varphi$, for all $w\in W$, implies $\overline{\square\varphi}=f_\square(\overline{\varphi})=f_\bot$. Thus, $\mathcal{M}$ is a $L5$-model. Now we let $\gamma\colon V\rightarrow M$ be the assignment $x\mapsto\overline{x}$. By induction on formulas, $\gamma(\varphi)=\overline{\varphi}$, for any $\varphi\in Fm$. Then $(\mathcal{M},\gamma)\vDash\varphi\Leftrightarrow\gamma(\varphi)=\overline{\varphi}\in\mathit{TRUE}\Leftrightarrow (w_T,g)\vDash\varphi$. Q.E.D.

\begin{definition}\label{280}
Let $\Phi\cup\{\varphi\}$ be a set of formulas. The relation of logical consequence w.r.t. Kripke semantics is defined as follows. $\Phi\Vdash^{Kr}_{L5}\varphi :\Leftrightarrow$ for every L5-frame $(W,R)$, every assignment $\gamma\colon V\rightarrow Pow(W)$ and \textit{maximal} world $w_T\in W$, $(w_T,\gamma)\vDash\Phi$ implies $(w_T,\gamma)\vDash\varphi$.
\end{definition}

So for Kripke semantics we have a pointwise (locally) defined consequence relation which only considers the \textit{maximal} points of a given frame. It follows by the definitions that if $(W,R)$ is a frame with maximal world $w_T$, $g$ is an assignment and $\varphi,\psi$ are formulas, then
\begin{equation}\label{600}
(w_T,g)\vDash\varphi\equiv\psi\Leftrightarrow\text{ for all }w\in W: (w,g)\vDash\varphi\text{ iff }(w,g)\vDash\psi.
\end{equation}
Recall that in modal logic, a \textit{proposition} is usually regarded as a set of possible worlds. Then \eqref{600} says that $\varphi\equiv\psi$ is true iff $\varphi$ and $\psi$ are satisfied at exactly the same worlds iff $\varphi$ and $\psi$ denote the same proposition. That is, $\varphi\equiv\psi$ actually stands for \textit{propositional identity}. In this sense, \eqref{600} is the analogue to \eqref{500} in terms of possible worlds semantics.

\begin{corollary}[Completeness w.r.t. Kripke semantics]\label{300}
Let $\Phi\cup\{\varphi\}$ be a set of formulas. Then 
\begin{equation*}
\Phi\vdash_{L5}\varphi \Leftrightarrow \Phi\Vdash_{L5}\varphi \Leftrightarrow \Phi\Vdash^{Kr}_{L5}\varphi.
\end{equation*}
\end{corollary} 

\paragraph*{Proof.}
The first equivalence is Theorem \ref{170} above, which can be proved in the same way as the corresponding completeness result for $L$ presented in \cite{lewjlc2}. The second equivalence follows by Theorems \ref{250} and \ref{270}. Q.E.D.

\section{The parametrized logics $L5(I)$}

In the following, we consider \textit{parametrized} versions of logic $L5$. Let $I$ be any intermediate logic. That is, $I$ results from $IPC$ by adding some axiom schemes that correspond to theorems of $CPC$. We write $\Phi\vdash_I\varphi$ if there is a derivation of $\varphi$ from $\Phi$ in $I$. By $L5(I)$ we denote the logic which results from $L5$ by considering in item (i) of the definition of $L5$ all theorems of $I$ instead of only those of $IPC$. In particular, $L5=L5(IPC)$. The notion of derivation $\Phi\vdash_{L5(I)}\varphi$ in $L5(I)$ is defined as usual.

We saw in \cite{lewjlc2} that $L5$ can be seen as a classical modal logic for the reasoning about intuitionistic truth, i.e. provability. Analogously, $L5(I)$ is a logic for the reasoning about truth in the sense of $I$. In the limit case $I=CPC$, the modal operator then becomes a predicate for classical truth in logic $L5(I)$ itself:

\begin{lemma}\label{400}
Let $I=CPC$. Then for all $\varphi\in Fm$, $\vdash_{L5(I)}\varphi\leftrightarrow\square\varphi$.
\end{lemma}

\paragraph*{Proof.}
The formula $\square\varphi\rightarrow\varphi$ is an axiom. We show that $\varphi\rightarrow\square\varphi$ is a theorem of $L5(CPC)$. First, observe that \textit{tertium non datur} $\varphi\vee\neg\varphi$ is not only a theorem but also an axiom of $L5(CPC)$. By rule AN and the axiom of the disjunction property, $\square\varphi\vee\square\neg\varphi$ is a theorem. Then $\varphi\rightarrow\square\varphi\vee\square\neg\varphi$ is a theorem. By axiom (ii) and $CPC$, $(\varphi\wedge\square\neg\varphi)\rightarrow (\varphi\wedge\neg\varphi)$. Thus, $\neg (\varphi\wedge\square\neg\varphi)$ is a theorem. By $CPC$, that is equivalent to $\neg\varphi\vee\neg\square\neg\varphi$ and to $\varphi\rightarrow\neg\square\neg\varphi$. Then we have $\varphi\rightarrow ((\square\varphi\vee\square\neg\varphi) \wedge\neg\square\neg\varphi)$. By distributivity, $\varphi\rightarrow ((\square\varphi\wedge\neg\square\neg\varphi)\vee(\square\neg\varphi\wedge\neg\square\neg\varphi))$ which is equivalent to $\varphi\rightarrow (\square\varphi\wedge\neg\square\neg\varphi)$. Of course, $(\square\varphi\wedge\neg\square\neg\varphi)\rightarrow\square\varphi$ is derivable. By transitivity, $\varphi\rightarrow\square\varphi$ is a theorem. Q.E.D.\\

Let $M(L5(I))$ be the class of those $L5$-models which evaluate all theorems of $I$ to the top element, under all assignments. That is, $\mathcal{M}\in M(L5(I))$ iff $\mathcal{M}$ is a $L5$-model and $\gamma(\varphi)=f_\top$ for all $I$-theorems $\varphi\in Fm_0$ and for all $\gamma\in M^V$.\footnote{By SP, such a model evaluates not only $I$-theorems to the top element but also any formula $\varphi\in Fm$ which has the form of an $I$-theorem and possibly contains the modal operator $\square$.} We refer to the elements of $M(L5(I))$ as $L5(I)$-models. Analogously, we define a $L5(I)$-frame as a $L5$-frame with the property that $(w_B,g)\vDash\varphi$ for all theorems $\varphi$ of $I$ and all assignments $g$, where $w_B$ is the bottom world. For a given set of formulas $\Phi\cup\{\varphi\}$, we write $\Phi\Vdash_{L5(I)}\varphi$ if $(\mathcal{M},\gamma)\vDash\Phi$ implies $(\mathcal{M},\gamma)\vDash\varphi$, for all $\mathcal{M}\in M(L5(I))$ and all assignments $\gamma$ in $\mathcal{M}$. Analogously, we define $\Phi\Vdash_{L5(I)}^{Kr}\varphi$ as in Definition \ref{280}, but with $L5(I)$-frames instead of all $L5$-frames. Now observe that Theorem \ref{250} assigns to each $L5(I)$-model $\mathcal{M}$ a $L5(I)$-frame $(W,R)$. In fact, if $\varphi$ is a theorem of $I$, then $(\mathcal{M},\gamma)\vDash\square\varphi$. By Theorem \ref{250}, $(w_T,g)\vDash\square\varphi$. This means $(w_B,g)\vDash\varphi$. On the other hand, Theorem \ref{270} assigns to each $L5(I)$-frame $(W,R)$ a $L5(I)$-model $\mathcal{M}$. For if $\varphi$ is an $I$-theorem, then $(w_B,g)\vDash\varphi$. Thus, $(w_T,g)\vDash\square\varphi$, with maximal world $w_T$. Then by Theorem \ref{270}, $(\mathcal{M},\gamma)\vDash\square\varphi$. That is, $\gamma(\varphi)=f_\top$. We conclude:

\begin{corollary}\label{490}
For any set $\Phi\cup\{\varphi\}\subseteq Fm$:
\begin{equation*}
\Phi\Vdash_{L5(I)}\varphi \Leftrightarrow \Phi\Vdash^{Kr}_{L5(I)}\varphi.
\end{equation*}
\end{corollary} 

By induction on derivations, we may prove soundness of $L5(I)$ w.r.t. the semantics generated by the class of all $L5(I)$-models: 
\begin{equation}\label{990}
\Phi\vdash_{L5(I)}\varphi \Rightarrow \Phi\Vdash_{L5(I)}\varphi.
\end{equation}
Now suppose $\Phi\cup\{\varphi\}\subseteq Fm_0$, i.e. we are given propositional formulas. As in [Lemma 2.3 \cite{lewjlc2}], one shows by induction on derivations that $\Phi\vdash_I\varphi$ implies $\square\Phi\vdash_{L5(I)}\square\varphi$. On the other hand, if $\Phi\nvdash_I\varphi$, then, in a similar way as in the proof of [Theorem 5.1 \cite{lewjlc2}] (in fact, it suffices to replace $IPC$ with $I$ in that proof), we may find a model $\mathcal{M}\in M(L5(I))$ and an assignment $\gamma$ such that $(\mathcal{M},\gamma)\vDash\square\Phi$ and $(\mathcal{M},\gamma)\nvDash\square\varphi$. That is, $\square\Phi\nVdash_{L5(I)}\square\varphi$. By soundness, $\square\Phi\nvdash_{L5(I)}\square\varphi$. We have established the following two results for \textit{propositional} $\Phi\cup\{\varphi\}\subseteq Fm_0$:

\begin{equation}\label{1000}
\begin{split}
&\Phi\vdash_{I}\varphi\Leftrightarrow\square\Phi\Vdash_{L5(I)}\square\varphi\\
&\Phi\vdash_{I}\varphi\Leftrightarrow\square\Phi\vdash_{L5(I)}\square\varphi.
\end{split}
\end{equation}

Note that we did not need completeness to establish \eqref{1000}. Nevertheless, completeness of $L5(I)$ can be shown in a similar way as completeness of $L$ \cite{lewjlc2}. Thus, the converse of \eqref{990} above holds true, too. The second statement of \eqref{1000} is a generalization of the Main Theorem of \cite{lewjlc2} with $I$ instead of $IPC$ and $L5(I)$ instead of $L$. What does that result mean? It is clear that $\Phi\vdash_{CLC}\varphi$ implies $\Phi\vdash_{L5(I)}\varphi$ (recall that $L5(I)$ contains all classical theorems). Now suppose $\Phi\vdash_{L5(I)}\varphi$, for propositional $\Phi\cup\{\varphi\}$. By soundness, $\Phi\Vdash_{L5(I)}\varphi$. In particular, if the two-element Boolean algebra (which, of course, is a $L5(I)$-model) satisfies $\Phi$, under a given assignment, then it also satisfies $\varphi$. This means that $\varphi$ follows from $\Phi$ in $CPC$. Thus, $\Phi\vdash_{CPC}\varphi\Leftrightarrow\Phi\vdash_{L5(I)}\varphi$, for propositional formulas $\Phi\cup\{\varphi\}$. This, together with the second statement of \eqref{1000}, shows that $L5(I)$ can be seen as a combination of intermediate logic $I$ and $CPC$. In particular, $L5(I)$ is a conservative extension of $CPC$, and $L5(I)$ contains a copy of $I$ in the following sense: $\vdash_I\varphi$ $\Leftrightarrow$ $\vdash_{L5(I)}\square\varphi$, for propositional $\varphi\in Fm_0$.  \\

Recall that $\square\varphi\leftrightarrow (\varphi\equiv \top)$ is a theorem of $L$ and of $L5(I)$. For a set of formulas $\Phi$, we write $\Phi\equiv\top$ for the set of equations $\{\psi\equiv\top\mid\psi\in\Phi\}$. Then the first statement of \eqref{1000} can be expressed in the following way. For propositional $\Phi\cup\{\varphi\}$:
\begin{equation}\label{1010}
\Phi\vdash_{I}\varphi\Leftrightarrow (\Phi\equiv\top) \Vdash_{L5(I)} (\varphi\equiv\top).
\end{equation}

Before we discuss \eqref{1010}, we define the \textit{reduct} of a $L5$-model (or a $L$-model) as the underlying Heyting algebra. Since a model has an ultrafilter, its reduct is a non-trivial Heyting algebra, i.e. it has at least two elements $f_\bot\neq f_\top$. Moreover, the reduct is a Heyting algebra with disjunction property DP. On the other hand, one easily shows that any non-trivial Heyting algebra with DP expands to a $L5$-model. In fact, the resulting $L5$-model only depends on the actual choice of the designated ultrafilter $\mathit{TRUE}$. Note that the operation $f_\square$ is uniquely determined in a $L5$-model.\\

These considerations show that we can interpret \eqref{1010} in the following way. 

\noindent $\Phi\vdash_{I}\varphi$ iff for the reduct of any $L5(I)$-model and any assignment, if all formulas of $\Phi$ denote the top element, then $\varphi$ denotes the top element. 

That is, we get a concept of logical consequence defined in terms of Heyting algebras. This corresponds to the usual notion of logical consequence w.r.t. algebraic semantics for $IPC$ found in the literature. However, whereas the usual notion involves \textit{all} Heyting algebras, we see here that it is enough to consider Heyting algebras with DP. This observation will play a crucial role in the next section.

How can we interpret \eqref{1010} under Kripke semantics? By Corollary \ref{490}, we have $\Phi\vdash_{I}\varphi\Leftrightarrow (\Phi\equiv\top) \Vdash_{L5(I)}^{Kr} (\varphi\equiv\top)$. By definition, this means that whenever $w_T$ is a maximal world of a $L5(I)$-frame and $g$ is any assignment, then $(w_T,g)\vDash\Phi\equiv\top$ implies $(w_T,g)\vDash\varphi\equiv\top$. But $(w_T,g)\vDash\Phi\equiv\top$ means $(w_B,g)\vDash\psi\leftrightarrow\top$ for all $\psi\in\Phi$, where $w_B$ is the bottom world. $(w_B,g)\vDash\psi\leftrightarrow\top$ implies $(w_B,g)\vDash\psi$. Consequently, we may express \eqref{1010} in the following way.
\begin{equation}\label{1020}
\Phi\vdash_{I}\varphi\Leftrightarrow\text{ if } (w_B,g)\vDash\Phi\text{ then }(w_B,g)\vDash\varphi,
\end{equation}
whenever $w_B$ is the bottom world of a $L5(I)$-frame and $g$ is any assignment.\\

Note that we have now two frame-based locally defined logical consequence relations. The first one, based on Definition \ref{280}, models logical consequence in the parametrized modal logics $L5(I)$ and involves only the \textit{maximal} worlds of a given frame. The second one, given in \eqref{1020}, models consequence in intermediate logics $I$ and involves the \textit{smallest} world of a given frame. This is the usual definition of logical consequence based on intuitionistic Kripke frames.

\section{A simple method for determining algebraic and Kripke semantics of some intermediate logics}

The results from the preceding section give rise to a simple method for determining algebraic and Kripke-style semantics of some specific intermediate logics. The method essentially relies on the fact that it suffices to work with Heyting algebras having DP. If intermediate logic $I$ is given as $I = IPC + \varphi_1 + ... + \varphi_n$ with \textit{disjunctive} and not too complicated formulas $\varphi_i$, then we may hope that our method is applicable. In the following, we illustrate the method discussing some specific examples with $n=1$. We obtain simple proofs of already known completeness results.

\subsection{The Logic of Here-and-There $HT$}

The Logic of Here-and-There ($HT$) was originally introduced by Heyting \cite{hey} as a three-valued logic for the purpose of showing that $IPC$ is strictly weaker than $CPC$. It reappeared in \cite{god} where G\"odel proved that $IPC$ cannot be characterized by a finite matrix of truth values. G\"odel also showed that $HT$ is the strongest intermediate logic weaker than $CPC$. Semantically, $HT$ can also be described by Heyting algebras with at most three elements and by Kripke frames with at most two worlds (the world of ``here" and the world of ``there"). $HT$ is also known as Smetanich Logic. The importance of $HT$ for logic programming under the stable semantics paradigma \cite{gellif} was discovered by D. Pearce \cite{pea1, pea2}. Moreover, results of Lifschitz, Pearce and Valverde \cite{lifpeaval} show that $HT$ can be seen as an adequate logic for reasoning with logic programs. Two logic programs are said to be equivalent if they have the same answer sets (stable models). This concept of equivalence, however, is not context independent. In \cite{lifpeaval}, two logic programs $P_1$ and $P_2$ are said to be \textit{strongly} equivalent if for any program $P$, the programs $P_1\cup P$ and $P_2\cup P$ are equivalent. In this sense, the concept of strong equivalence is independent of the actual context in which logic programs are embedded. The authors show that two logic programs are strongly equivalent iff they are equivalent as formulas in $HT$. Observe now that by \eqref{1000} above, for any propositional formulas $\varphi,\psi$:
\begin{equation*}
\vdash_{HT}\varphi\leftrightarrow\psi \text{ if and only if } \vdash_{L5(HT)}\varphi\equiv\psi.
\end{equation*}
That is, the relation of propositional identity $\varphi\equiv\psi$, which is defined as strict equivalence $\square(\varphi\leftrightarrow\psi)$ in the sense of Lewis' modal logics, reads as strong equivalence of corresponding logic programs. Note that the above discussed context independence of \textit{strong} equivalence, defined in \cite{lifpeaval}, is in some sense expressed by theorem SP: $(\varphi\equiv\psi)\rightarrow (\chi[x:=\varphi]\equiv\chi[x:=\psi])$, our representation of the principle of \textit{Indiscernibility of Identicals}, shortly discussed in the introductory part. \\

Hosoi \cite{hos} proved that Kripke semantics of $HT$ can be axiomatized by $IPC+[\varphi\vee(\varphi\rightarrow\psi)\vee\neg\psi]$. Recently, a more direct proof was found by Harrison et al. \cite{harlifpearval}.

In the following, we illustrate our method deriving algebraic and Kripke semantics directly from Hosoi's axiomatization. This results in a further proof of Hosoi's theorem. 

By \eqref{1010}, $HT$ is sound and complete w.r.t. the class of Heyting algebras which are reducts of $L5(HT)$-models. We will characterize those reducts by their algebraic structure. By \eqref{1010}, $\Vdash_{L5(HT)} (x\vee(x\rightarrow y)\vee\neg x)\equiv\top$. Let $\mathcal{H}$ be the reduct of a $L5(HT)$-model and suppose that $m,m'$ are elements of $\mathcal{H}$ distinct from the top and the bottom. We consider an assignment $\gamma$ with $\gamma(x)=m$ and $\gamma(y)=m'$. Then $\gamma(x\vee(x\rightarrow y)\vee\neg y)=\gamma(\top)=f_\top$. By DP, $\gamma(x)=f_\top$ or $\gamma(x\rightarrow y)=f_\top$ or $\gamma(\neg y)=f_\top$. By hypothesis, $\gamma(x)\neq f_\top$ and $\gamma(\neg y)=\gamma(y\rightarrow\bot)\neq f_\top$. Hence, $\gamma(x\rightarrow y)=f_\rightarrow(m,m')=f_\top$. That is, $m\le m'$. Now we consider an assignment $\beta$ with $\beta(x)=m'$ and $\beta(y)=m$ and conclude in a similar way that $m'\le m$. Hence, $m=m'$ and $\mathcal{H}$ has exactly three elements: $m$, $f_\top$ and $f_\bot$. One also easily checks that a reduct may have only two elements: $f_\top$ and $f_\bot$. We have shown that the reduct of any $L5(HT)$-model is a Heyting algebra with at most three elements. Now suppose we are given a non-trivial Heyting algebra with at most three elements. Note that such an algebra is a linearly ordered: $f_\bot \le m \le f_\top$. Then it is clear that Hosoi's axiom $x\vee (x\rightarrow y)\vee\neg y$ is satisfied, under all assignments. Hence, the reducts of $L5(HT)$-models are precisely the non-trivial Heyting algebras with at most three elements, and $HT$ is sound and complete w.r.t. that class of algebras. 

There is exactly one Heyting algebra with three elements, and the unique Heyting algebra with two-elements is the two-element Boolean algebra (up to isomorphisms). Obviously, the Boolean algebra has only one (prime) filter, and the three-element Heyting algebra has exactly two (prime) filters which are linearly ordered by inclusion. By Theorem \ref{250}, this results in frames with at most two worlds $w_B,w_T$ (possibly $w_B=w_T$). On the other hand, suppose we are given a frame with at most two worlds $w_T, w_B$. Then one verifies that $\varphi\approx\psi\Leftrightarrow (w_B,g)\vDash\varphi\leftrightarrow\psi$ defines an equivalence relation with at most three classes $\overline{\top}$, $\overline{\bot}$ and $\overline{\varphi}$, where $\varphi$ is any formula false at the bottom world and true at the top world. According to the proof of Theorem \ref{270}, this results in a $L5$-model with at most three elements, i.e. a $L5(HT)$-model. Hence, $HT$ is sound and complete w.r.t. Kripke semantics generated by frames with at most two worlds. Q.E.D.

\subsection{G\"odel-Dummett Logic $G$}

M. Dummett \cite{dum} considers the logic $IPC+[(\varphi\rightarrow\psi)\vee(\psi\rightarrow\varphi)]$ and shows its completeness w.r.t. algebraic semantics given by all linearly ordered Heyting algebras. The logic is known as G\"odel-Dummett Logic $G$ because of its relations to G\"odel's $n$-valued logics studied in \cite{god}. J. v. Plato \cite{pla} observed that that logic was introduced by T. Skolem already in 1913. P. H\'ajek \cite{haj} studies $G$ as one of the important Fuzzy Logic systems which are given as extensions of H\'ajek's basic logic $BL$.

A relatively simple proof of Dummett's original completeness theorem is found by A. Horn \cite{hor}. Horn's proof is based on the fact that a Heyting algebra $\mathcal{H}$ validates Dummett's axiom iff $\mathcal{H}$ is a subalgebra of a direct product of linearly ordered Heyting algebras. A similar proof, in terms of $BL$-algebras, is contained in \cite{haj}. In the following, we prove Dummett's theorem with our method.\\

By \eqref{1010}, $G$ is sound and complete w.r.t. the class of Heyting algebras wich are reducts of $L5(G)$-models. Our goal is to characterize those algebras by their specific structure. By \eqref{1010}, $\Vdash_{L5(G)}((x\rightarrow y)\vee (y\rightarrow x))\equiv\top$. Then for any given $L5(G)$-model and assignment $\gamma$, $\gamma((x\rightarrow y)\vee (y\rightarrow x))=f_\top$. By DP, $\gamma(x\rightarrow y)=f_\top$ or $\gamma(y\rightarrow x)=f_\top$. That is, $\gamma(x)\le \gamma(y)$ or $\gamma(y)\le \gamma(x)$. This holds for all assignments. Thus, the universe of the model is linearly ordered. We have proved that the reduct of every $L5(G)$-model is a linearly ordered Heyting algebra. On the other hand, it is clear that every linearly ordered Heyting algebra evaluates Dummett's axiom $(x\rightarrow y)\vee (y\rightarrow x)$ to the top element, under all assignments. We conclude that the class of $L5(G)$ reducts is exactly the class of all non-trivial linearly ordered Heyting algebras. Hence, $G$ is sound and complete w.r.t. the semantics generated by that class of algebras. 

Note that logic $HT$ axiomatizes a special class of linearly ordered Heyting algebras, namely those with at most three elements. Consequently, $G$ is a sublogic of $HT$. What can be said about the prime filters of a linearly ordered Heyting algebra? We may consider such an universe as the closed interval $[f_\bot,f_\top]$ which is linearly ordered by the underlying lattice ordering. The supremum (infimum) of two elements $m,m'$ equals $m$ or $m'$. Then it is clear that the filters are precisely the unions of closed intervals $[m,f_\top]$ with $m > f_\bot$. In particular, all filters are prime, and they are linearly ordered by inclusion (observe that the unique ultrafilter is the set $(f_\bot,f_\top]=\bigcup_{m\neq f_\bot}[m,f_\top]$). By Theorem \ref{250}, this results in linearly ordered frames. Now suppose we are given a frame $(W,R)$ which is a linear ordering. Again, we consider the equivalence relation $\varphi\approx\psi\Leftrightarrow (w_B,g)\vDash\varphi\leftrightarrow\psi$ from the proof of Theorem \ref{270}. For two elements $\overline{\varphi}$, $\overline{\psi}$ of the resulting $L5$-model, we have $\overline{\varphi}\le\overline{\psi}\Leftrightarrow f_\rightarrow(\overline{\varphi},\overline{\psi})=f_\top\Leftrightarrow\overline{\varphi\rightarrow\psi}=\overline{\top}\Leftrightarrow (w_B,g)\vDash\varphi\rightarrow\psi$. Since the worlds are linearly ordered and $w_B$ is the bottom world, one easily checks that $(w_B,g)\nvDash\varphi\rightarrow\psi$ implies $(w_B,g)\vDash\psi\rightarrow\varphi$. Then by the above equivalences, $\overline{\varphi}\nleq\overline{\psi}$ implies $\overline{\psi}\le\overline{\varphi}$. That is, the resulting $L5$-model is linearly ordered, i.e. it is a $L5(G)$-model. Consequently, Kripke semantics of logic $G$ is given by the class of linearly ordered frames. Q.E.D.

\subsection{Jankov Logic $KC$}

The logic axiomatized by $IPC+\neg\varphi\vee\neg\neg\varphi$ was introduced by V. A. Jankov \cite{jan} and is known as Jankov Logic, $KC$ or the Logic of the Weak Law of the Excluded Middle. Jankov proved its soundness and completeness w.r.t. finite rooted Kripke frames with a single maximal world. D. de Jongh and L. Hendriks \cite{jonhen} showed that $KC$ is the weakest intermediate logic for which strongly equivalent logic programs, in a language allowing negations, are logically equivalent. In the following, we show how algebraic and Kripke-style semantics of $KC$ derives from Jankov's axiomatization using our general method. 

By \eqref{1010}, $KC$ is sound and complete w.r.t. the class of Heyting algebras which are reducts of $L5(KC)$-models. We aim at a characterization of those algebras. For any reduct of a $L5(KC)$-model and any assignment $\gamma$, we have $\gamma(\neg x\vee\neg\neg x)=f_\top$. By DP, $\gamma(\neg x)=f_\top$ or $\gamma(\neg\neg x)=f_\top$. This is equivalent to the condition: 
\begin{equation}\label{1450}
\gamma(x)=f_\bot\text{ or }\gamma(x\rightarrow\bot)=f_\rightarrow(\gamma(x),f_\bot)=f_\bot.
\end{equation} 
Recall that the relative pseudo-complement $f_\rightarrow(\gamma(x),f_\bot)$ of $\gamma(x)$ w.r.t. $f_\bot$ is the \textit{greatest} element $m$ such that $f_\wedge(\gamma(x),m)\le f_\bot$. Then, with \eqref{1450}, $\gamma(x)> f_\bot$ implies $f_\rightarrow(\gamma(x),f_\bot)=f_\bot$ implies $f_\wedge(\gamma(x),m') > f_\bot$, for all $m' > f_\bot$. This holds for all assignments $\gamma$. We conclude that the reduct of any $L5(KC)$-model is a non-trivial Heyting algebra with DP and the following specific property. For all elements $m,m'$:
\begin{equation}\label{1500}
m > f_\bot\text{ and }m' > f_\bot \Rightarrow f_\wedge(m,m') > f_\bot.
\end{equation}

Let us refer to such Heyting algebras as $KC$-algebras. In order to characterize the class of reducts of $L5(KC)$-models as precisely the class of $KC$-algebras, it remains to show that every $KC$-algebra is the reduct of a $L5(KC)$-model, i.e. evaluates the formula $\neg x\vee\neg\neg x$ to the top element, under any assignment. Suppose we are given a $KC$-algebra and an assignment $\gamma$ with $\gamma(\neg x\vee\neg\neg x)\neq f_\top$. Then $\gamma(\neg x)=\gamma(x\rightarrow \bot)=f_\rightarrow(\gamma(x),f_\bot)\neq f_\top$ and $\gamma(\neg\neg x)=\gamma(\neg x\rightarrow \bot)=f_\rightarrow(\gamma(\neg x),f_\bot))\neq f_\top$. Thus, $\gamma(x) > f_\bot$ and $\gamma(\neg x)=f_\rightarrow(\gamma(x),f_\bot) > f_\bot$. However, $f_\wedge(\gamma(x),f_\rightarrow(\gamma(x),f_\bot))=f_\bot$, as in every Heyting algebra. This contradicts the property of a $KC$-algebra, condition \eqref{1500} above. Hence, $\gamma(\neg x\vee\neg\neg x) = f_\top$. We have proved that the reducts of $L5(KC)$-algebras are precisely the $KC$ algebras. Hence, Jankov logic is sound and complete w.r.t. the semantics given by the class of $KC$-algebras. Note that the models of G\"odel-Dummett Logic $G$ are special $KC$-algebras. Hence, $KC \subseteq G \subseteq HT$.  

Let us specify the corresponding Kripke semantics. We claim that each $KC$-algebra has exactly one ultrafilter. Suppose there are two ultrafilters $U\neq U'$. Then there is some $m\in U\smallsetminus U'$. By Lemma \ref{200}, $f_\neg(m):=f_\rightarrow(m,f_\bot)\in U'$. Because $m$ and $f_\neg(m)$ belong to filters, they are greater than the bottom element. However, their infimum equals the bottom. This contradicts the specific property \eqref{1500} of a $KC$-algebra. Thus, a $KC$-algebra has exactly one ultrafilter. By Theorem \ref{250}, this results in frames with a single maximal world. Now suppose we are given a frame with a single maximal world $w_T$. For a given assignment $g$, we consider again the equivalence class $\approx$ on $Fm$ defined in the proof of Theorem \ref{270}. We must show that the resulting $L5$-model is a $L5(KC)$-model, i.e. has the property \eqref{1500} of a $KC$-algebra. So let $\overline{\varphi}\neq\overline{\bot}$ and $\overline{\psi}\neq\overline{\bot}$ be two elements greater than the bottom. Since neither $\varphi\approx\bot$ nor $\psi\approx\bot$, there are worlds $w$ and $w'$ with $(w,g)\vDash\varphi$ and $(w',g)\vDash\psi$. Both worlds must access the same maximal world because there is only one, namely $w_T$. Then, by monotonicity, $(w_T,g)\vDash\varphi\wedge\psi$. That is, $f_\wedge(\overline{\varphi},\overline{\psi})=\overline{\varphi\wedge\psi}\neq\overline{\bot}=f_\bot$, and \eqref{1500} is fulfilled. Hence, the resulting Heyting algebra is a $KC$-algebra. We conclude that Kripke semantics for $KC$ is given by all frames with a single maximal world. It is known that $IPC$ is complete w.r.t. the class of all \textit{finite} rooted Kripke models. Since $IPC\subseteq KC$, it suffices to consider \textit{finite} frames with a single maximal world as Kripke semantics for Jankov Logic. Q.E.D.

\end{document}